\documentclass[a4paper,12pt]{article}

\usepackage[utf8x]{inputenc}
\usepackage{graphicx}
\usepackage{overcite}
\usepackage{amsmath}
\usepackage{amssymb}
\usepackage{float}
\usepackage{rotating}
\usepackage{longtable}
\usepackage{scrextend}
\usepackage[font=small,labelfont=bf]{caption}
\usepackage{url}
\usepackage{bm} % Bold maths symbols (e.g., \bm{\phi})

\usepackage{setspace}

\newcommand{\onlinecite}[1]{\hspace{-1 ex} \nocite{#1}\citenum{#1}}

\title{\bf Error-Controlled Exploration of Chemical Reaction Networks with Gaussian Processes}
\author{Gregor N.\ Simm and Markus Reiher\thanks{corresponding author: markus.reiher@phys.chem.ethz.ch}
\vspace{10 mm}\\ ETH Z\"urich, Laboratory of Physical Chemistry, \\ Vladimir-Prelog-Weg 2, 8093 Z\"urich, Switzerland.
}

\begin{document}

\maketitle

\begin{center}
{\bf Abstract}
\end{center}
{\small
For a theoretical understanding of the reactivity of complex chemical systems,
relative energies of stationary points on potential energy hypersurfaces need to be calculated to high accuracy.
Due to the large number of intermediates present in all but the simplest chemical processes, approximate quantum chemical methods are required that allow for fast evaluations of the relative energies but
at the expense of accuracy.
Despite the plethora of benchmark studies, the accuracy of a quantum chemical method is often difficult to assess.
Moreover, a significant improvement of a method's accuracy (e.g., through reparameterization or systematic model extension) is rarely possible.
Here, we present a new approach that allows for the systematic, problem-oriented, and rolling improvement of quantum chemical results through the application of Gaussian processes.
Due to its Bayesian nature, reliable error estimates are provided for each prediction.
A reference method of high accuracy can be employed, if the uncertainty associated with a particular calculation is above a given threshold.
The new data point is then added to a growing data set in order to continuously improve the model, and as a result, all subsequent predictions.
Previous predictions are validated by the updated model to ensure that uncertainties remain within the given confidence bound, which we call
backtracking.
We demonstrate our approach at the example of a complex chemical reaction network.
}

\setstretch{1.5}

\section{Introduction}

The accurate description of chemical processes requires elucidation of a reaction network comprising all relevant intermediates and elementary reactions.
For a kinetic analysis, the thermodynamic properties of all intermediates and transition states in this network need to be determined to high accuracy.
While state-of-the-art quantum chemical calculations can yield highly accurate results even for large systems,\cite{Claeyssens2006}
they are computationally expensive and, therefore, restricted to a limited number of elementary steps.
For this reason, density functional theory (DFT) remains to be the method of choice, despite its shortcomings with respect to accuracy and systematic improvability.
The known existence of the exact exchange-correlation functional\cite{Hohenberg1964,Lieb1983} and numerical demonstration\cite{Mardirossian2017}
of rung-by-rung accuracy of approximate functionals across
Jacob's ladder\cite{Perdew2001} have nurtured the hope to eventually arrive at approximate functionals of sufficiently high quality.
However, popular hybrid density functionals struggle to reproduce experimental ligand dissociation energies of large transition-metal complexes (see, e.g., refs.\ \cite{Niu2000,Schultz2005,Furche2006,Riley2007,Jiang2012,Weymuth2014,Weymuth2015,Liu2015,Husch2018,Ma2018}).

The accuracy of approximate exchange--correlation density functionals is often assessed through benchmark studies.
While many extensive data sets exist, such as the ones proposed by Pople,\cite{Pople1989,Curtiss1991,Curtiss1997,Curtiss2000}
Truhlar,\cite{Lynch2003a,Lynch2003b,Lynch2003,Zhao2004,Schultz2005a,Schultz2005,Zhao2005,Zhao2005a,Zhao2008,Zhao2009}
and Grimme,\cite{Korth2009,Goerigk2010,Goerigk2011} studies have shown that the accuracy of density functionals can be strongly system-dependent.\cite{Curtiss1997,Curtiss2000,Salomon2002,Zhao2004,Curtiss2005,Riley2007,Weymuth2014}
Even if accurate reference data for given molecular structures were available, one could not assume the error of a DFT result to be transferable to the close vicinity of that particular region of chemical space.\cite{Cohen2008,Weymuth2014,Pernot2015,Liu2015,Simm2016,Husch2018}
Finally, even if one had some upper bound on the error of a calculated property, this bound would be so large that subsequent analyses, such as kinetic studies, would be rendered meaningless.\cite{Simm2017}
If, however, an error estimate of each result was known, the value of any approximate DFT approach would be dramatically increased as it would flag those results to be considered with caution.
An assigned uncertainty would allow one to judge whether conclusions drawn from the data are valid or not.

A few methods have been developed that aim at providing systematic error estimates for individual DFT results.
In 2005, Sethna, N{\o}rskov, Jacobsen, and co-workers presented an error estimation scheme based on Bayesian statistics \cite{Mortensen2005} (see also refs.~\onlinecite{Brown2003,Frederiksen2004,Petzold2012}).
Instead of focusing on only the best-fit parameters in a density functional, an ensemble of parameters was drawn from a conditional distribution over parameters
by which a mean and a variance could be assigned to each computational result.\cite{Wellendorff2012,Wellendorff2014,Pandey2015}
While the developed functionals were parametrized on a wide range of reference data sets, the issue of transferability remained.
In addition, the reliability of the error estimates was limited.\cite{Pernot2017}
Recently, Zabaras and co-workers\cite{Aldegunde2016} developed a new exchange-correlation functional to predict bulk properties of transition metals and monovalent semiconductors.
Furthermore, Vlachos and co-workers successfully applied Bayesian statistics to DFT reaction rates on surfaces.\cite{Sutton2016}

In 2016, we presented a Bayesian framework for DFT error estimation based on the ideas of Sethna and co-workers\cite{Mortensen2005}
that allows for error estimation of calculated molecular properties.
By system-focused reparameterization of a long-range corrected hybrid (LCH) density functional, we obtained an accurate functional that yielded reliable confidence intervals
for reaction energies in a specific reaction network.
Unfortunately, the accuracy of the functional and the error estimates were limited by the flexibility of the LCH functional chosen as a starting point. The error estimates did not improve systematically with the size of the data set.
In addition, the process of reparameterizing the functional, which is necessary for this type of uncertainty quantification when new reference data points are incorporated,
is cumbersome and slow because quantum chemical calculations must be repeated for the whole data set.

Over the last years, many studies on the application of statistical learning to chemistry have been published,
with applications ranging from electronic structure predictions (e.g., refs.~\onlinecite{Behler2007,Balabin2009,Bartok2010,Rupp2012,Montavon2013,Bartok2013a,Hansen2013,Schutt2014,Dral2015,De2016,Faber2017,Chmiela2017,Schutt2017a,Bartok2017})
to applications in force-field development (e.g., refs.~\onlinecite{Behler2008,Handley2010,Botu2015a,Hansen2015,Shen2016,Li2017,Bleiziffer2018,Chmiela2018}),
materials discovery (e.g., refs.~\onlinecite{Gomez-Bombarelli2016,Zhou2017,Altae-Tran2017,Liu2017a,Segler2018}),
and reaction prediction.\cite{Kayala2011,Sadowski2016,Wei2016,Raccuglia2016,Segler2017,Jin2017,Fooshee2018,Bradshaw2018,Ahneman2018}
For recent reviews on the applications of machine learning in chemistry, see refs.~\onlinecite{Rupp2015a} and \onlinecite{Ramakrishnan2017}.

De Vita, Csányi, and co-workers presented a scheme that combines \textit{ab initio} calculation and machine learning for molecular dynamics simulations.\cite{Csanyi2004,Li2015,Glielmo2017,Glielmo2018}
Forces on atoms are either predicted by Bayesian inference or, if necessary, computed by on-the-fly quantum-mechanical calculations and added to a growing
machine learning database.\cite{Li2015}
However, this approach requires a considerable data set size to be accurate.
So far, their approach was applied to the simulation of metal solids but not to molecular systems.

In 2017, N{\o}rskov, Bligaard, and co-workers employed Gaussian processes (GPs) to construct a surrogate model on the fly to efficiently study surface reaction networks involving hydrocarbons.\cite{Ulissi2017}
The surrogate model is iteratively used to predict the rate-limiting reaction step to be calculated explicitly with DFT.
In their study, extended connectivity fingerprints based on graph representations of molecules are applied to represent adsorbed species.
However, if the uncertainties provided by the GP are high, then reference calculations are not automatically performed to improve the model.
Therefore, the construction of the reference data set is not directly guided by the GPs' predictions.
Finally, their approach was applied to study surface chemistry, for which more accurate \textit{ab initio} approaches,
typically coupled cluster methods, are not applied on a routine basis.

Despite continuous advances, most machine learning approaches are unsuitable for the study of chemical reactivity.
Training data sets, which are required for the learning process of the statistical model, are commonly assembled by drawing from a predefined pool of chemical species.
This approach would only be applicable to the exploration of a chemical system if the specific species had been known before
(which cannot be achieved as these species are the result of the exploration process).
By contrast, structure discovery through exploration requires system-focused uncertainty quantification in order to be reliable.\cite{Simm2016}
While some machine learning methods provide error estimates for such system-focused, rolling approaches, in most studies applying statistical learning to investigate chemical systems
the focus is placed on the prediction accuracy (e.g., refs.~\onlinecite{Behler2007,Balabin2009,Bartok2010,Rupp2012,Montavon2013,Bartok2013a,Hansen2013,Schutt2014,Dral2015,De2016,Faber2017,Chmiela2017,Schutt2017a,Bartok2017}).
In molecular applications\cite{Peterson2017,Janet2017},
confidence intervals are not exploited to define structures for which reference data should be calculated in a rolling fashion.

Here, we present an approach that addresses the aforementioned limitations.
A GP is employed to predict properties for species encountered during the exploration.
Due to the Bayesian nature of GPs, confidence intervals are provided for each prediction, and the uncertainty attached to each result can be assessed.
If the prediction confidence is below a certain threshold, the result will be flagged and accurate quantum chemical methods can be employed to obtain
more reliable data for the cases singled out by the GP.
Subsequently, this data point is added to the data set, the GP is retrained to incorporate the new data point, and its predictions improve.
In this way, the GP can be systematically improved; a larger data set will result in more accurate predictions.
Naturally, the size of this data set depends on the desired accuracy.
However, the focus of this study is not on the machine learning method and its accuracy
but instead on its feature to provide error estimates that allow us to structure reference data and to determine where more reference data are needed.

We demonstrate our approach with the example of a model reaction network consisting of isomers of $\text{C}_7\text{H}_{10}\text{O}_2$ stoichiometry
and consider DFT and semiempirical approaches as approximate models whose reliability is to be assessed in a system-focused way.
We emphasize that our approach is applicable to any kind of electronic structure model, ranging from semiempirical and tight-binding models to
multiconfigurational approaches with multireference perturbation theory, provided that results of higher accuracy are available
for several reference points selected by our algorithm.

\section{Theory}

\subsection{Gaussian Process Regression -- Overview}

GPs have been extensively studied by the machine learning community. They are rooted in a sophisticated and consistent theory combined with
computational feasibility.\cite{Rasmussen2006}
In chemistry, however, GPs are fairly new, and therefore, a short overview is given here.
We refer the reader to ref.~\onlinecite{Rasmussen2006} for a more detailed derivation.

Supervised learning is the problem of learning input to output mappings from a training data set.
We define the training data set containing $N$ observations as $\mathcal{D} = \{(\textbf{x}_i, y_i)| i = 1, \ldots, N\}$, where $\mathbf{x}$ is the input and $y$ the output.
From $\mathcal{D}$, we aim for learning the underlying function $f$ to make predictions for an unseen input $\mathbf{x_\ast}$, i.e., input that is not in $\mathcal{D}$.
Because no function that reproduces the training data is equally valid, it is necessary to make assumptions about the characteristics of $f$.
With a GP, which is a stochastic \textit{process} describing distributions over functions,\cite{Rasmussen2006}
one includes all possible functions and assigns weights to these functions depending on how likely they are to model the underlying function.

By defining a \textit{prior} distribution, we encode our prior belief on the function that we are trying to model.
The prior distribution over functions includes not only the mean and point-wise variance over the functions at a certain point $\mathbf{x}$
but also how smooth these functions are.
The latter is encoded in the covariance function or \textit{kernel}, which determines how rapidly the functions should change based on a change in the input $\mathbf{x}$.
The task of \textit{learning} is finding the optimal values for the parameters in the model.
The \textit{posterior} distribution is the result of combining the prior and the knowledge that we get from $\mathcal{D}$.
With a trained GP, one can make predictions on unseen input.
Due to its Bayesian nature, an error estimate, indicating the model's confidence in the prediction, is provided for each prediction.
Finally, the GP is systematically improvable, i.e., predictions and their error estimates improve with data set size.

\subsection{Gaussian Process Regression -- Brief Derivation}

Let us consider a simple linear regression model with Gaussian noise
\begin{equation}\label{eq:linear_model}
  f(\mathbf{x}) = \bm{\phi}(\mathbf{x})^\intercal \mathbf{w}, \qquad\qquad y = f(\mathbf{x}) + \varepsilon,
\end{equation}
where $\mathbf{x}$ is a $D$-dimensional input vector, $\mathbf{w}$ is a vector of parameters, and $y$ is the observed target value.
The function $\bm{\phi}(\mathbf{x})$ maps a $D$-dimensional input vector to a $D'$-dimensional feature space.
Moreover, we assume that the observed target value $y$ differs from $f$ by some noise $\varepsilon$,
which obeys an independent and identically distributed Gaussian distribution $\mathcal{N}$ with a mean and variance $\sigma_n^2$
\begin{equation}\label{eq:noise}
  \varepsilon \sim \mathcal{N}(0, \sigma_n^2).
\end{equation}
Furthermore, as our prior, we place a zero-mean Gaussian with covariance matrix $\Sigma_p$ on the weights
\begin{equation}\label{eq:weights}
  \mathbf{w} \sim \mathcal{N}(0, \Sigma_p).
\end{equation}

Following Bayes' rule, the posterior distribution reads
\begin{equation}\label{eq:bayes_rule}
	p(\mathbf{w} | X, \mathbf{y}) = \frac{p(\mathbf{y} | X, \mathbf{w}) \, p(\mathbf{w} | X)}{p(\mathbf{y} | X)},
\end{equation}
where $X = \{\mathbf{x}_i | i = 1,\ldots,N\}$ and $\mathbf{y} = [y_1, \ldots, y_N]^\intercal$.
In eq.~\eqref{eq:bayes_rule}, the marginal likelihood, $p(\mathbf{y} | X)$, is independent of the weights and can be calculated according to
\begin{equation}\label{eq:marginal_likelihood}
	p(\mathbf{y} | X) = \int p(\mathbf{y} | X, \mathbf{w}) p(\mathbf{w}) \, d\mathbf{w}.
\end{equation}

For some unseen $\mathbf{x_\ast}$, the probability distribution of $f(\mathbf{x_\ast})$ is given by the following expression:
\begin{equation}\label{eq:predict}
  p(f_\ast | \mathbf{x_\ast}, X, \mathbf{y}) = \int p(f_\ast | \mathbf{x_\ast}, \mathbf{w}) p(\mathbf{w} | X, \mathbf{y}) \, d\mathbf{w}.
\end{equation}
This can be shown to be\cite{Rasmussen2006}
\begin{equation}\label{eq:predict2}
\begin{split}
  p(f_\ast | \mathbf{x_\ast}, X, \mathbf{y})
  = \mathcal{N} \Big( &\bm{\phi}_\ast^\intercal \Sigma_p \mathbf{\Phi} (\mathbf{\Phi^\intercal} \Sigma_p \mathbf{\Phi} + \sigma_n^2I)^{-1} \mathbf{y}, \\
  &\bm{\phi}_\ast^\intercal \Sigma_p \bm{\phi}_\ast - \bm{\phi}_\ast^\intercal \Sigma_p \mathbf{\Phi} (\mathbf{\Phi^\intercal} \Sigma_p \mathbf{\Phi} + \sigma_n^2I)^{-1} \mathbf{\Phi^\intercal} \Sigma_p \bm{\phi}_\ast \Big),
\end{split}
\end{equation}
where $\bm{\phi}_\ast = \bm{\phi}(\mathbf{x_\ast})$
and $\mathbf{\Phi} = \mathbf{\Phi}(X)$ is the column-wise aggregation of $\bm{\phi}(\mathbf{x})$ for all inputs in $\mathcal{D}$.
In eq.~\eqref{eq:predict2}, the feature space always enters in the form of $\bm{\phi}(\mathbf{x})^\intercal \Sigma_p \bm{\phi}(\mathbf{x'})$,
where $\mathbf{x}$ and $\mathbf{x'}$ are in either the training or test set.
It is useful to define the \textit{covariance function} or \textit{kernel} $k(\mathbf{x}, \mathbf{x'}) = \bm{\phi}(\mathbf{x})^\intercal \Sigma_p \bm{\phi}(\mathbf{x'})$
and the corresponding kernel matrix $K(X, X') = \mathbf{\Phi}(X)^\intercal \Sigma_p \mathbf{\Phi}(X')$.
Because the covariance matrix $\Sigma_p$ is positive semidefinite, we can define $\Sigma^{1/2}$ so that $\Big(\Sigma_p^{1/2}\Big)^2 = \Sigma_p$.
Therefore, we can write $\bm{\phi}(\mathbf{x})^\intercal \Sigma_p \bm{\phi}(\mathbf{x'})$ as an inner product $\langle \mathbf{\psi(x)}, \mathbf{\psi(x')}\rangle$, where $\mathbf{\psi(x)} = \Sigma_p^{1/2} \bm{\phi}(\mathbf{x})$.
This is also known as the \textit{kernel trick}, which allows one to circumvent the explicit representation of the function $\bm{\phi}$ in eq.~\eqref{eq:linear_model}.
Conveniently, on the basis of Mercer's theorem,\cite{Mercer1909} it suffices to verify that $k(\mathbf{x}, \mathbf{x'})$ satisfies Mercer's condition.
For a more elaborate explanation, see section 4.3 in ref.~\onlinecite{Rasmussen2006}.
Finally, the key predictive equations for a GP regression are:\cite{Rasmussen2006}
\begin{equation}\label{eq:predict_distr}
\mathbf{f_\ast} | X, \mathbf{y}, X_\ast \sim \mathcal{N}\left( \mathbf{\overline{f}_\ast}, \text{cov}(\mathbf{f_\ast})\right),
\end{equation}
where
\begin{equation}\label{eq:predict_mean}
\mathbf{\overline{f}_\ast} \triangleq \mathbb{E}[\mathbf{f_\ast} | X, \mathbf{y}, X_\ast] = K(X_\ast, X)[K(X, X) + \sigma_n^2 I]^{-1} \mathbf{y}
\end{equation}
and
\begin{equation}\label{eq:predict_var}
\text{cov}(\mathbf{f_\ast}) = K(X_\ast, X_\ast) - K(X_\ast, X)[K(X,X) + \sigma_n^2 I]^{-1} K(X, X_\ast).
\end{equation}

A GP trained on $\mathcal{D}$ to make predictions on $f$ can be employed to model functions such as:
\begin{equation}
  g(\mathbf{x}, \mathbf{x}') = f(\mathbf{x}) - f(\mathbf{x}').
\end{equation}
The prediction mean can be readily obtained from the individual prediction means
\begin{equation}\label{eq:mean_diff}
  \overline{g}(\mathbf{x}, \mathbf{x}') = \overline{f}(\mathbf{x}) - \overline{f}(\mathbf{x}')
\end{equation}
and the prediction uncertainty can be estimated employing the individual variances and covariance $\text{cov}(f(\mathbf{x}), f(\mathbf{x}'))$, which can be computed with eq.~\eqref{eq:predict_var}:
\begin{equation}\label{eq:cov_diff}
  \text{cov}(g(\mathbf{x}, \mathbf{x}')) = \text{cov}(f(\mathbf{x})) + \text{cov}(f(\mathbf{x}')) - 2 \, \text{cov}(f(\mathbf{x}), f(\mathbf{x}')).
\end{equation}

\subsection{Molecular Kernels}

From eqs.~\eqref{eq:predict_mean} and \eqref{eq:predict_var} it can be seen that
in order to be able to apply GPs to learn a molecular target $\mathcal{T}(\mathbf{x})$ (e.g., an enthalpy of atomization), the kernel $k(\mathbf{x}, \mathbf{x}')$ needs to be evaluated.
Here, $\mathbf{x}$ may be some point in chemical space, i.e., the atomic configuration, charge, and spin multiplicity.
The kernel should measure the similarity between two points in chemical space and satisfy invariance properties such as translations, rotations, and permutation of atoms of the same element.
The search for new kernels to encode physical invariances is a subject of active research.

If the target $\mathcal{T}(\mathbf{x})$ can be approximately decomposed as a sum of local contributions, the formulation of the kernel can be simplified:
\begin{equation}\label{eq:decomposition}
  \mathcal{T}(\mathbf{x}) = \sum_{\ell = 1}^{n} t(\tilde{x}_\ell),
\end{equation}
where $\ell$ is an atomic index, $n$ is the total number of atoms, and $\tilde{x}_\ell$ is a local atomic environment.
This approximation can be appropriate for properties such as the energy or molecular polarizability.\cite{Angyan1994}
Then, we can model $t(\tilde{x}_\ell)$ as a linear combination of abstract descriptors $\tilde{\bm{\phi}}(\tilde{x}_\ell)$ (see eq.~\eqref{eq:linear_model}):
\begin{equation}
  \hat{t}(\tilde{x}_\ell) = \tilde{\bm{\phi}}(\tilde{x}_\ell)^\intercal \mathbf{w}.
\end{equation}
In analogy to equation \eqref{eq:decomposition}, we obtain
\begin{equation}
  \hat{\mathcal{T}}(\mathbf{x}) = \sum_{\ell = 1}^{n} \tilde{\bm{\phi}}(\tilde{x}_\ell)^\intercal \mathbf{w} = \bm{\phi}(\mathbf{x})^\intercal \mathbf{w},
\end{equation}
where $\bm{\phi}(\mathbf{x}) = \sum_{\ell = 1}^{n} \tilde{\bm{\phi}}(\tilde{x}_\ell)$ so that we recover eq.~\eqref{eq:linear_model}.
One can see that the kernel $k(\mathbf{x}, \mathbf{x'})$ can be written as a sum of kernels acting on local atomic environments
\begin{equation}
  k(\mathbf{x}, \mathbf{x'}) = \bm{\phi}(\mathbf{x})^\intercal \Sigma_p \bm{\phi}(\mathbf{x'}) = \sum_{\ell = 1}^{n} \sum_{\ell' = 1}^{n'} \tilde{k}(\tilde{x}_\ell, \tilde{x}'_{\ell'}),
\end{equation}
where $\tilde{k}(\tilde{x}_\ell, \tilde{x}'_{\ell'}) = \tilde{\bm{\phi}}(\tilde{x}_\ell) \sum_p \tilde{\bm{\phi}}(\tilde{x}'_{\ell'})$.
There are many kernels developed to act on atomic environments $\tilde{k}(\tilde{x}_\ell, \tilde{x}'_{\ell'})$, such as the kernel developed by Behler and Parrinello\cite{Behler2007},
the Smooth Overlap of Atomic Potentials (SOAP)\cite{Bartok2013}, or the Graph Approximated Energy (GRAPE).\cite{Ferre2017}

\subsection{Error-Controlled Exploration Protocol}
\label{subsec:protocol}

In the exploration of a chemical reaction network, the data set $\mathcal{D}$ is not known beforehand and must be generated during the exploration
for a system-focused uncertainty quantification.
Naturally, the size of this data set should be related to the desired level of confidence with which the target $\mathcal{T}$ needs to be determined.
Our protocol starts with an initial training data set $\mathcal{D}$ of size $m > 0$ and the desired level of confidence given by the variance $\sigma^2_\text{thresh}$.
The initial data set consists of the first $m$ structures $s_{1:m} = \{ \mathbf{x}_1, ..., \mathbf{x}_m \}$ encountered during the exploration and the corresponding targets.
This is necessary to allow for reliable predictions by the learning algorithm.
However, it is critical that the initial training data set does not result in the model being overly confident.
Therefore, the optimal choice of $m$ depends on the chemical system and the exploration method.
For example, if $\mathcal{D}$ consisted of $m$ consecutive snapshots of a molecular dynamics trajectory,
$m$ should be chosen to be larger than if it contained largely different configurational isomers.
We also note that one could construct the initial data set by sampling the configuration space employing an inexpensive method
and, subsequently, applying a clustering algorithm (e.g., $k$-means clustering) so that $\mathcal{D}$ consists of the centroids of the $m$ clusters.

Subsequently, new structures $s_{m+1:N}$ (given by a list of structures here (see the Supporting Information) but constructed in a rolling fashion in practice) are encountered.
Each structure $\textbf{x}_i$ is fed to the GP, and a prediction mean $\bar{\mathcal{T}}(\textbf{x}_i)$ and a variance $\sigma^2_i$ are obtained.
If $\sigma^2_i$ is less than $\sigma^2_\text{thresh}$, the prediction confidence will be sufficiently high and the next structure will be attained.
If $\sigma^2_i$ is larger than $\sigma^2_\text{thresh}$, the prediction will be discarded and the target will be explicitly calculated
(e.g., with an electronic structure reference method) for that structure.
The newly obtained data point is added to $\mathcal{D}$ and the GP is retrained on the extended data set.
Naturally, there is a trade-off between confidence and computational effort.
If $\sigma^2_\text{thresh}$ is decreased, the prediction confidence will be required to be higher throughout the exploration.
This requires a larger data set and, hence, more reference calculations.
If, however, $\sigma^2_\text{thresh}$ is increased, fewer reference calculations are needed, but the overall prediction accuracy is lower.
Next, all predictions made before are repeated with the updated GP.
Through this process, which we refer to as \textit{backtracking}, we ensure that predictions on previously encountered structures
are still within the given confidence interval after the GP was updated.
Our error-controlled exploration protocol with backtracking can be summarized as:\\

%\begin{algorithm}
%\caption{Error-controlled exploration strategy.}
%\label{alg:exploration}
%\begin{algorithmic}
%  \INPUT $\mathcal{D} = \{(\textbf{x}_i, \mathcal{T}(\textbf{x}_i))\}_{i=1}^{m}$, $s_{m+1:N}$, $\sigma^2_\text{thresh}$
{\bf Input:} $\mathcal{D} = \{(\textbf{x}_i, \mathcal{T}(\textbf{x}_i))\}_{i=1}^{m}$, $s_{m+1:N}$, $\sigma^2_\text{thresh}$\\[-3ex]

%  \For{$ i \gets m+1, N $}
  ~~{\bf for} {$ i \gets m+1, N $} {\bf do}\\[-3ex]

%    \State $\bar{\mathcal{T}}(\textbf{x}_i) \gets \mathbb{E}_{GP}[\mathcal{T}(\textbf{x}_i) | \mathcal{D}, \textbf{x}_i]$
    ~~~~~~$\bar{\mathcal{T}}(\textbf{x}_i) \gets \mathbb{E}_{GP}[\mathcal{T}(\textbf{x}_i) | \mathcal{D}, \textbf{x}_i]$\\[-3ex]

%    \State $\sigma^2_i \gets \mathbb{V}_{GP}[\mathcal{T}(\textbf{x}_i) | \mathcal{D}, \textbf{x}_i]$
    ~~~~~~$\sigma^2_i \gets \mathbb{V}_{GP}[\mathcal{T}(\textbf{x}_i) | \mathcal{D}, \textbf{x}_i]$\\[-3ex]

%    \If{$ \sigma^2_i > \sigma^2_\text{thresh}$}
    ~~~~~~{\bf if} {$ \sigma^2_i > \sigma^2_\text{thresh}$} {\bf then}\\[-3ex]

%      \State add $(\textbf{x}_i, \mathcal{T}(\textbf{x}_i))$ to $\mathcal{D}$
      ~~~~~~~~~~add $(\textbf{x}_i, \mathcal{T}(\textbf{x}_i))$ to $\mathcal{D}$\\[-3ex]

%      \State update GP and backtrack (i.e., check $x_{j<i}$)
      ~~~~~~~~~~update GP and backtrack (i.e., check $x_{j<i}$)\\[-3ex]

%    \EndIf
%  \EndFor

%  \Return $\mathcal{D}$
  {\bf return} $\mathcal{D}$
%\end{algorithmic}
%\end{algorithm}

\section{Results}

\subsection{Model System}

We demonstrate our error-controlled exploration strategy with the example of a subset of the GDB-9 database\cite{Ramakrishnan2014}
consisting of three-dimensional molecular structures of 6095 constitutional isomers of the $\text{C}_7\text{H}_{10}\text{O}_2$ stoichiometry.
We chose this database in order to adhere to a publicly available data set that promotes reproducibility and comparability of new
algorithms such as the one proposed in section \ref{subsec:protocol}.

\begin{figure}
  \centering
  \includegraphics[width=\textwidth]{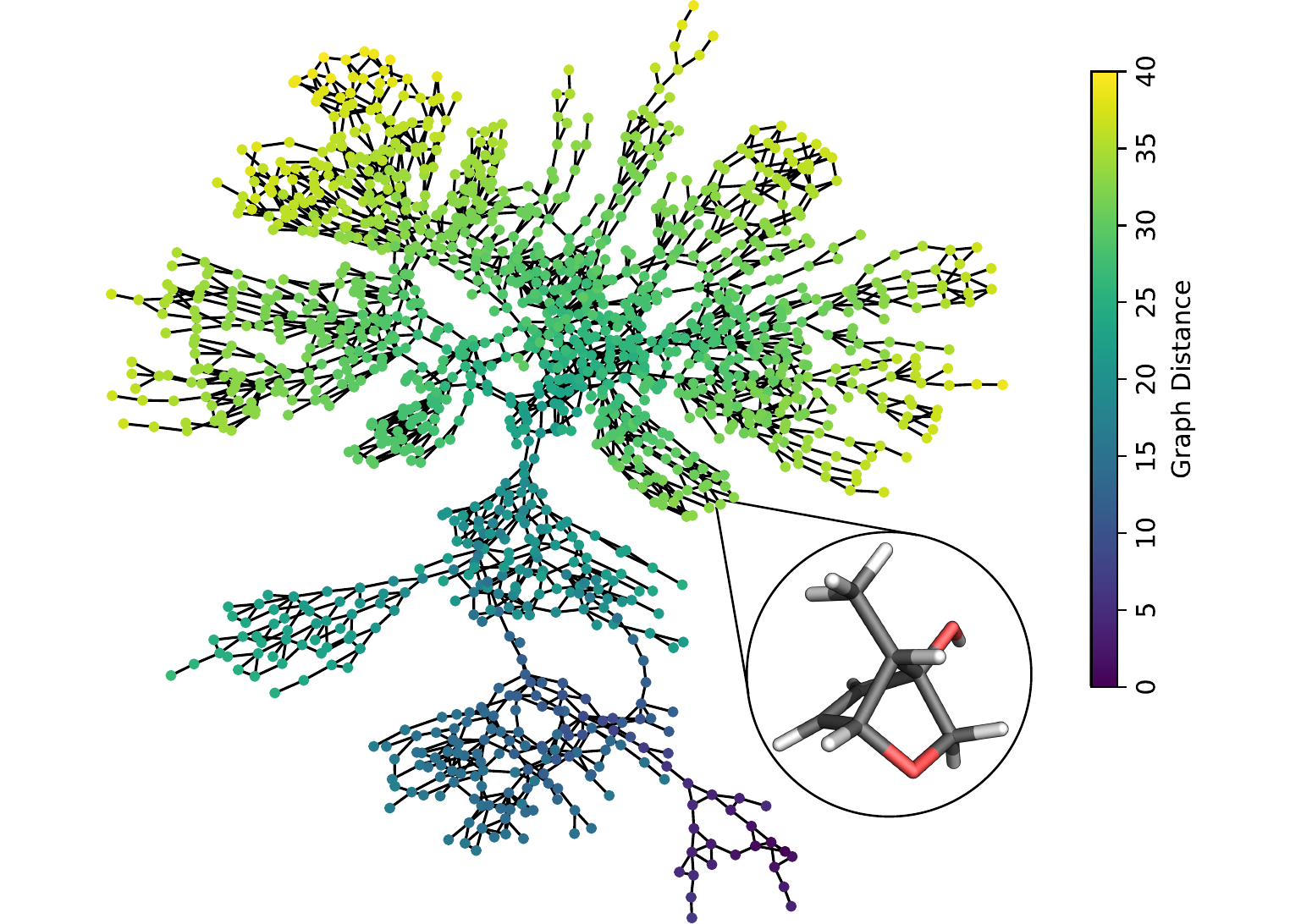}
  \caption{
  Reaction network considered in this study.
  Nodes represent three-dimensional molecular structures of constitutional isomers of the $\text{C}_7\text{H}_{10}\text{O}_2$ stoichiometry.
  Edges are drawn between two nodes if there is a transformation rule interconverting their molecular graphs.
  A node's color represents its graph distance to a (randomly chosen) node in the network.
  }
  \label{fig:wave}
\end{figure}

We constructed a graph in which nodes represent items in this data set.
Edges are placed between two nodes if their molecular graphs can be interconverted by at least one rule from a set of transformation rules.
These rules describe reactions commonly found in organic chemistry including nucleophilic addition and substitution,
isomerization, and cycloaddition reactions (see the Supporting Information for details).
The application of these rules divided this graph into multiple strongly connected subgraphs, the largest of which contained 1494 nodes.
This subgraph will serve as an artificial exploration network for the rest of this article and is provided in the Supporting Information.
The exploration network is shown in Fig.~\ref{fig:wave}.
The color of each node represents the graph distance to some randomly chosen node in the network, i.e., the number of edges in the shortest path connecting them.

We calculate the SOAP kernel\cite{Bartok2013} $k(\mathbf{x}, \mathbf{x}')$ for every pair of structures in the data set.
This measure of molecular similarity is suitable for a special class of molecular structures that we consider in this work: stable intermediates.
In fact, many electronic structure methods ranging from Kohn--Sham DFT to single-reference coupled cluster models have been developed
for this special type of stationary points on the Born--Oppenheimer potential energy hypersurface (PES).
It is well-known that many of them will fail for dissociation processes (examples are wrong asymptotes of coupled cluster calculations
and the Hartree--Fock dissociation error).
Clearly, considering also structures away from stable intermediates would require an extension of the descriptor chosen for this work.
However, such extensions are rather straightforward to define.
Consider, for example, a multidimensional descriptor that also considers electronic structure information such as the gap between
the highest occupied molecular orbital and the lowest unoccupied molecular orbital;
see also the work of Kulik and co-workers.\cite{Janet2017a,Janet2017,Janet2018}
Such an extension of the kernel would also improve its ability to capture long-range effects.

A special and important class of stationary points on the PES next to that of stable intermediates are transition-state structures,
i.e., first-order saddle points on the PES.
We would need to consider these structures in order to transgress the thermodynamic view of reaction networks and to approach kinetic modeling.
Whereas this is beyond the scope of the present work, we note in passing that apart from the option to explicitly include information on the electronic structure of a given molecular structure
(which would also allow one to consider different charge and spin states),
we may treat transition-state structures as a new class of structures characterized by the fact that an electronically excited state is generally closer in energy than that is the case for stable intermediate.
One may, therefore, keep intermediates and transition states (and species of different charge or spin multiplicity) in separate data sets
in order to best account for these different types of electronic structures (e.g., closed-shell ground-state minima, ground-state bond-activated structures with
a tendency for multiconfigurational nature, neutral vs.\ excess-charge species, and so forth).

If a set of intermediates on different PESs (but with the same charge and spin multiplicity) are encountered during the exploration,
the smallest collection of atoms from which every molecule in the set can be constructed can be assembled.
Then, upon comparison of two structures $x$ and $x'$ from this set with the kernel $k(\mathbf{x}, \mathbf{x}')$,
the atoms that are not needed to form either of the two would still be part of the comparison
but in the form of idealized ``isolated'' species.\cite{De2016}
In this way, all comparisons between structures from this set are on equal footing.

\subsection{Learning and Predictions}

Calculating a thermodynamic property $P^\text{ref} (\mathbf{x})$ (e.g., the standard enthalpy of atomization) with accurate methods, such as G4MP2,\cite{Curtiss2007} is computationally demanding.
Statistical learning can be employed to improve a result of computationally (comparatively) inexpensive quantum chemical methods, $P^\text{base} (\mathbf{x})$,
by predicting the error of a method with respect to some accurate reference result:
\begin{equation}
  \Delta P^\text{ref}_\text{base} (\mathbf{x}) = P^\text{ref} (\mathbf{x}) - P^\text{base} (\mathbf{x}).
\end{equation}
This strategy is often referred to as $\Delta$-machine learning.\cite{Ramakrishnan2015a}
It is based on the idea that inexpensive quantum chemical methods are able to describe a significant portion of the underlying physics (e.g., nuclear repulsion)
but fail to capture more complex phenomena such as electron correlation.
It is these effects that are then learned in a $\Delta$-machine learning approach.
By design, $\Delta$-machine learning approaches require the evaluation of the inexpensive $P^\text{base}$ to arrive at the desired property.

\begin{figure}
  \centering
  \includegraphics[width=0.5\textwidth]{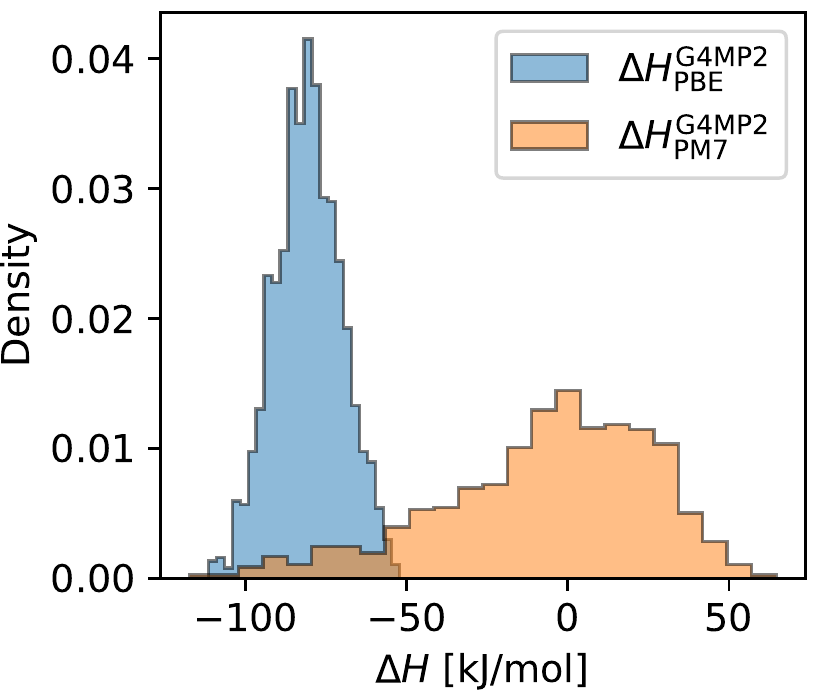}
  \caption{
  Distributions of $\Delta H^\text{G4MP2}_\text{PBE}$ and $\Delta H^\text{G4MP2}_\text{PM7}$ for the data set.
  }
  \label{fig:deltas}
\end{figure}

In this work, we apply the $\Delta$-machine learning approach by learning the difference in the calculated standard enthalpy of atomization between G4MP2
and the density-functional approach with PBE\cite{Perdew1996a} ($\Delta H^\text{G4MP2}_\text{PBE}$) as well as G4MP2 and the semiempirical
model PM7\cite{Stewart2012} ($\Delta H^\text{G4MP2}_\text{PM7}$).
We emphasize that the choice of inexpensive (here, PBE and PM7) and reference (here, G4MP2) method is to a certain degree arbitrary,
and other choices work as well for our protocol (provided that the reference method has been demonstrated to be more accurate than the inexpensive
models for the data set under consideration).
The distributions of $\Delta H^\text{G4MP2}_\text{PBE}$ and $\Delta H^\text{G4MP2}_\text{PM7}$ in the data set are shown in Fig.~\ref{fig:deltas}
(see the Computational Methodology for details).
Due to the more approximate nature of the semiempirical PM7 method compared to the PBE density functional,
the distribution of $\Delta H^\text{G4MP2}_\text{PM7}$ is much wider than that of $\Delta H^\text{G4MP2}_\text{PBE}$.

\begin{figure}
  \centering
  \includegraphics[width=0.5\textwidth]{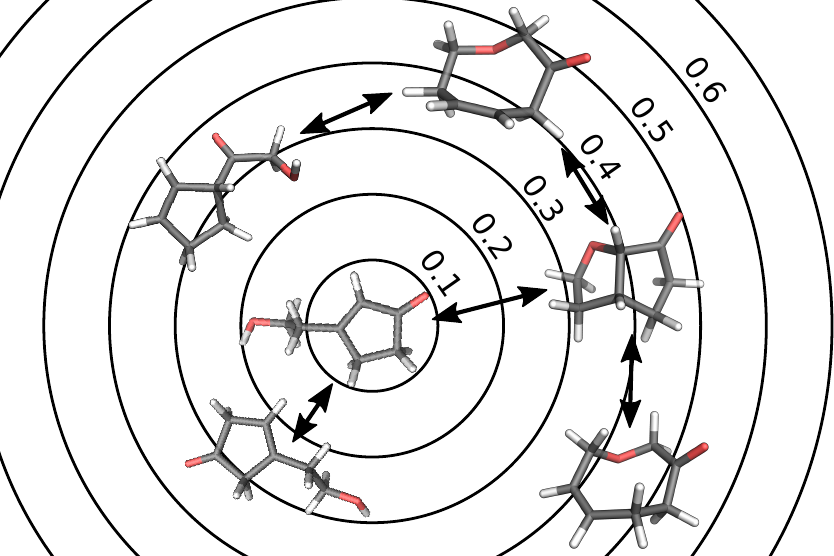}
  \caption{
  Illustration of the distance metric in eq.~\eqref{eq:distance} introduced by the kernel at the example of a reaction subnetwork.
  The contour lines represent the distance $d(\mathbf{x}, \mathbf{x}')$ between the reactant in the center ($\mathbf{x}$)
  and possible reaction products present in the data set ($\mathbf{x}'$).
  Double arrows are drawn between structures if there is a transformation rule interconverting their molecular graphs.
  }
  \label{fig:distances}
\end{figure}

We calculate the SOAP kernel\cite{Bartok2013} $k(\mathbf{x}, \mathbf{x}')$ for every pair of structures in the data set.
This kernel also provides a definition of the distance between two structures\cite{De2016}
\begin{equation}\label{eq:distance}
  d(\mathbf{x}, \mathbf{x}') = \sqrt{2 - 2 k(\mathbf{x}, \mathbf{x}')}.
\end{equation}
To illustrate the notion of distance in a reaction network,
a subnetwork of the whole reaction network is arranged according to $d(\mathbf{x}, \mathbf{x}')$ in Fig.~\ref{fig:distances}, where $\mathbf{x}$ is some reactant and $\mathbf{x}'$ a possible product.

For both targets separately, we trained a GP on randomly selected subsets of different size and employed the remaining structures as an out-of-sample validation set.
The GP's hyperparameters are optimized by maximizing the marginal likelihood.
For predictions on the validation set, we calculated the mean absolute error (MAE),
\begin{equation}
  \text{MAE} = \frac{1}{N} \sum_{i=1}^{N} | \bar{\mathcal{T}}(\mathbf{x}_i)- \mathcal{T}(\mathbf{x}_i) | ,
\end{equation}
and root-mean-square error (RMSE),
\begin{equation}
  \text{RMSE} = \sqrt{ \frac{1}{N} \sum_{i=1}^{N} \left( \bar{\mathcal{T}}(\mathbf{x}_i)- \mathcal{T}(\mathbf{x}_i) \right)^2 } ,
\end{equation}
where $N$ is the size of the out-of-sample validation set, $\bar{\mathcal{T}}(\mathbf{x}_i)$ the prediction mean, and $\mathcal{T}(\mathbf{x}_i)$ the target value.
To better assess the behavior of the GP, we also calculated the MAE (MAE$_\text{ref}$) and the RMSE (RMSE$_\text{ref}$) of a trivial statistical model
that simply predicts the mean of the training data set for every test input.
In addition, to guarantee the accuracy of the error estimates, we calculated the percentage of predictions $r_\text{cb}$ for which the target lies outside of the 95\% confidence band given by
$\bar{\mathcal{T}}(\mathbf{x}_i) \pm 2 \sigma (\mathbf{x}_i)$.
We repeated this process 25 times to ensure that the average of the above metrics converged.
The average properties are summarized in Table~\ref{tab:at}.
It can be seen that the prediction accuracy improves significantly with the size of the training data set.
When comparing the MAE and the RMSE to the MAE$_\text{ref}$ and the RMSE$_\text{ref}$, respectively,
it becomes evident that the use of a GP can be justified only for training data set sizes of 200 and larger.
It can also be seen that the prediction error of $\Delta H^\text{G4MP2}_\text{PM7}$ is larger than that of $\Delta H^\text{G4MP2}_\text{PBE}$.
This can be explained by the approximate nature of the semiempirical PM7 method (see Fig.~2).
Nonetheless, the results suggest that the prediction error estimates are reliable as $r_\text{cb}$ is close to 5\% for all data set sizes and targets.

\begin{table}
\centering
\caption{
Mean absolute error (MAE), reference MAE (MAE$_\text{ref}$), root-mean-square error (RMSE), reference RMSE (RMSE$_\text{ref}$)
(in kJ/mol), and $r_\text{cb}$ of GP predictions on $\Delta H^\text{G4MP2}_\text{PBE}$ and $\Delta H^\text{G4MP2}_\text{PM7}$ for different training data set sizes.
}
\label{tab:at}
\begin{tabular}{llrrrrr}
\hline
Size & Target  &  MAE &  MAE$_\text{ref}$ &  RMSE &  RMSE$_\text{ref}$ &  $r_\text{cb}$ \\
\hline
50   & $\Delta H^\text{G4MP2}_\text{PBE}$ &   7.82 &      8.42 &   9.71 &      10.53 &          5.24 \\
     & $\Delta H^\text{G4MP2}_\text{PM7}$ &  21.61 &     26.24 &  27.86 &      33.13 &          6.40 \\
100  & $\Delta H^\text{G4MP2}_\text{PBE}$ &   7.30 &      8.42 &   9.03 &      10.53 &          4.53 \\
     & $\Delta H^\text{G4MP2}_\text{PM7}$ &  19.15 &     26.16 &  25.01 &      32.99 &          6.03 \\
200  & $\Delta H^\text{G4MP2}_\text{PBE}$ &   6.37 &      8.40 &   7.84 &      10.50 &          3.52 \\
     & $\Delta H^\text{G4MP2}_\text{PM7}$ &  15.71 &     26.12 &  21.06 &      32.97 &          6.48 \\
500  & $\Delta H^\text{G4MP2}_\text{PBE}$ &   4.42 &      8.39 &   5.45 &      10.48 &          3.83 \\
     & $\Delta H^\text{G4MP2}_\text{PM7}$ &   8.31 &     26.16 &  11.25 &      32.99 &          6.21 \\
1000 & $\Delta H^\text{G4MP2}_\text{PBE}$ &   2.90 &      8.37 &   3.64 &      10.45 &          4.26 \\
     & $\Delta H^\text{G4MP2}_\text{PM7}$ &   4.64 &     26.15 &   6.21 &      32.91 &          4.74 \\
\hline
\end{tabular}
\end{table}

For the study of chemical reactivity, not enthalpies of formation but (free) enthalpy differences between intermediates are usually of interest.
From a GP trained on a molecular target, predictions on differences with respect to that target between molecular structures are readily available through eqs.~\eqref{eq:mean_diff} and \eqref{eq:cov_diff}.
For both targets separately, we trained a GP on randomly selected subsets of different size and then predicted relative energies between the remaining structures.
This process was repeated 25 times to obtain converged means of the MAE, RMSE, and $r_\text{cb}$.
From the results shown in Table~\ref{tab:rel_enthalpies}, it can be seen that the MAE and the RMSE decrease rapidly with data set size; however,
the accuracy is lower than that of predictions on the standard enthalpy of atomization.
Nonetheless, $r_\text{cb}$ indicates that the error estimates remain reliable.

\begin{table}
\centering
\caption{
Mean absolute error (MAE), root-mean-square error (RMSE) (in kJ/mol), and $r_\text{cb}$ of predictions on differences in the standard enthalpy between molecular structures
from GPs trained on targets $\Delta H^\text{G4MP2}_\text{PBE}$ and $\Delta H^\text{G4MP2}_\text{PM7}$.
}
\label{tab:rel_enthalpies}
\begin{tabular}{llrrr}
\hline
Size & Target &       MAE        &     RMSE &  $r_\text{cb}$  \\
\hline
50   & $\Delta H^\text{G4MP2}_\text{PBE}$ &         10.96 &          13.67 &                   5.35 \\
     & $\Delta H^\text{G4MP2}_\text{PM7}$ &         30.69 &          39.11 &                   6.34 \\
100  & $\Delta H^\text{G4MP2}_\text{PBE}$ &         10.22 &          12.74 &                   4.91 \\
     & $\Delta H^\text{G4MP2}_\text{PM7}$ &         27.54 &          35.26 &                   5.56 \\
200  & $\Delta H^\text{G4MP2}_\text{PBE}$ &          8.88 &          11.07 &                   4.22 \\
     & $\Delta H^\text{G4MP2}_\text{PM7}$ &         22.95 &          29.75 &                   5.81 \\
500  & $\Delta H^\text{G4MP2}_\text{PBE}$ &          6.17 &           7.70 &                   4.37 \\
     & $\Delta H^\text{G4MP2}_\text{PM7}$ &         12.13 &          15.88 &                   5.96 \\
1000 & $\Delta H^\text{G4MP2}_\text{PBE}$ &          4.09 &           5.15 &                   4.53 \\
     & $\Delta H^\text{G4MP2}_\text{PM7}$ &          6.72 &           8.78 &                   5.36 \\
\hline
\end{tabular}
\end{table}

Hence, we demonstrated that GPs are capable of learning molecular properties of molecular structures with reliable error estimates.
Furthermore, relative molecular properties can be predicted with sufficient accuracy employing a statistical model
trained on individual molecular properties.

\subsection{Error-Controlled Exploration}

\begin{figure}
  \centering
  \includegraphics[width=\textwidth]{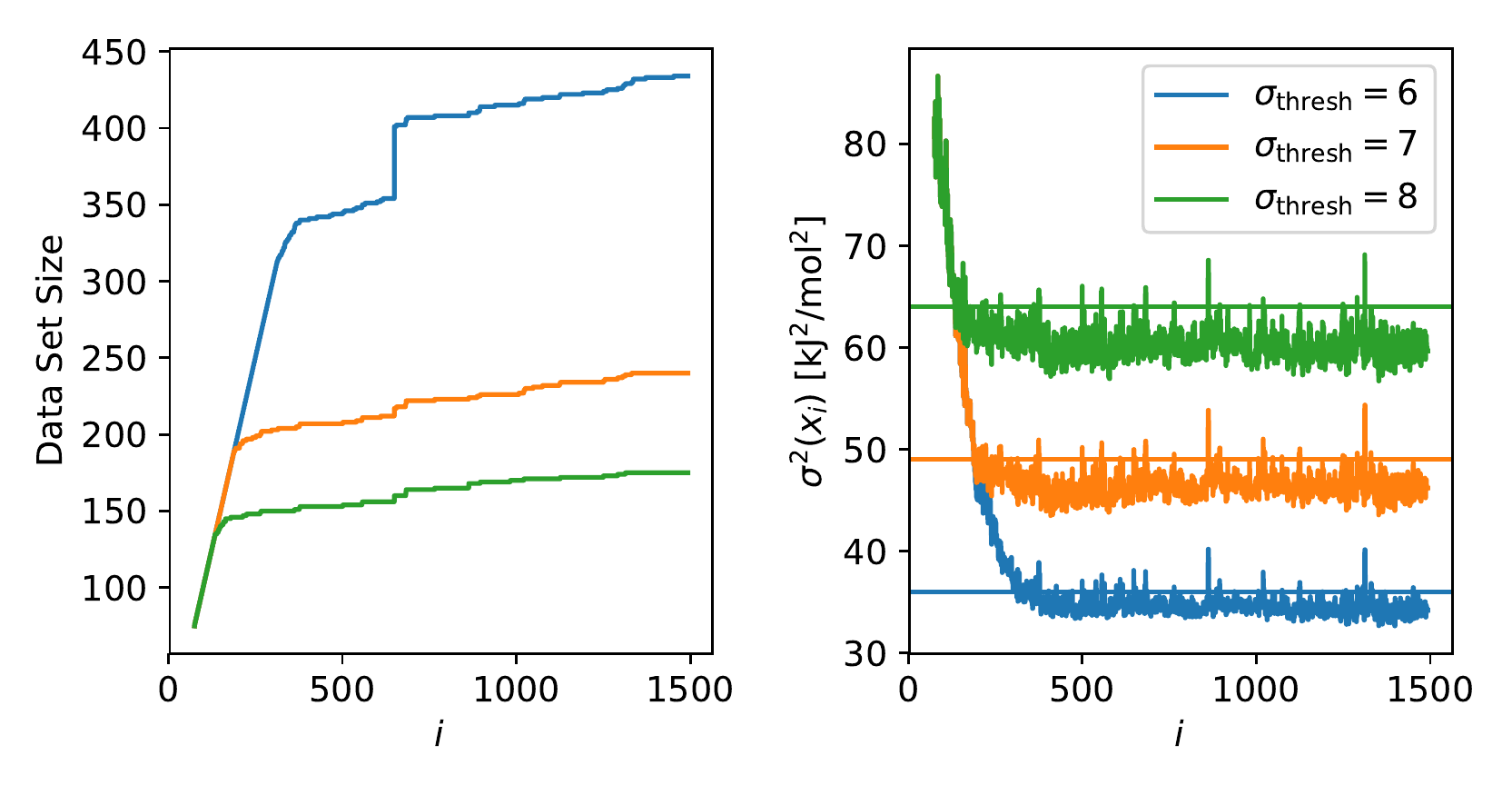}
  \caption{
  Size of the training data set (left) and prediction variance on the enthalpy of atomization (right) for the $i$th structure in an exploration
  employing the PBE functional and G4MP2 as the reference.
  }
  \label{fig:exploration_pbe}
\end{figure}

\begin{figure}
  \centering
  \includegraphics[width=\textwidth]{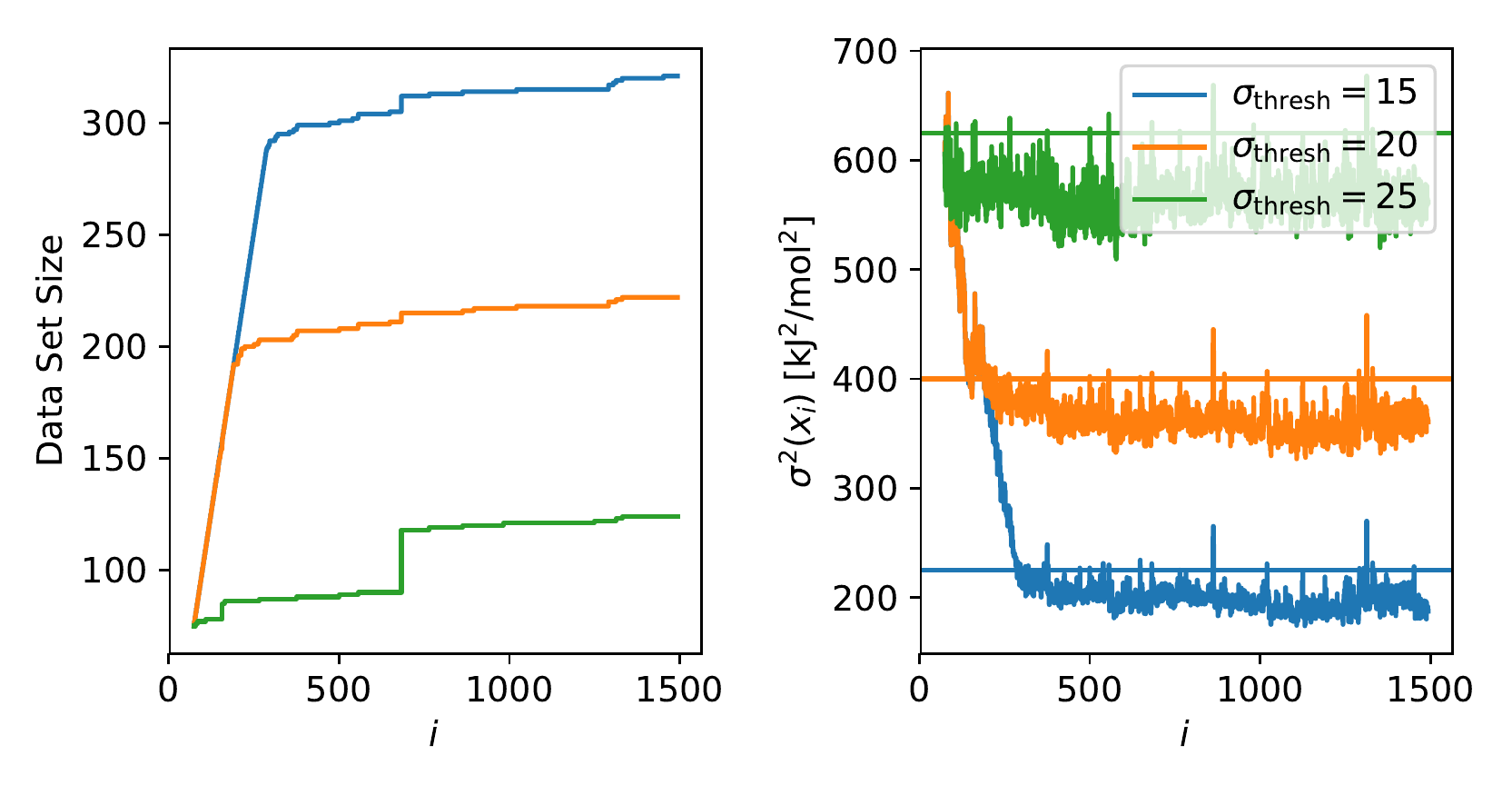}
  \caption{
  Size of the training data set (left) and prediction variance on the enthalpy of atomization (right) for the $i$th structure in an exploration
  employing PM7 and G4MP2 as the reference.
  }
  \label{fig:exploration_pm7}
\end{figure}

For the consecutive discovery of intermediates in the exploration of a chemical system, we generated sequences of nodes from our reaction network.
Whereas all nodes are already known in our example network, an actual exploration procedure would expand the network in a continuous fashion
(see refs.\ \onlinecite{Simm2017a} and \onlinecite{Bergeler2015}).
Starting from a random initial node in the reaction network,
the remaining nodes were visited in the order of their graph distance to the initial node (see Fig.~\ref{fig:wave}).
Nodes with the same graph distance were discovered in a random order.
Next, the error-controlled exploration strategy outlined in section \ref{subsec:protocol} was applied.
Here, the initial data set consisted of the first $m=75$ explored nodes.
The explorations were separately performed for the targets $\Delta H^\text{G4MP2}_\text{PBE}$ and $\Delta H^\text{G4MP2}_\text{PM7}$.
For each target, three different runs with different variance thresholds were carried out.
Results for the exploration with targets $\Delta H^\text{G4MP2}_\text{PBE}$ and $\Delta H^\text{G4MP2}_\text{PM7}$ (on the same sequence) are shown
in Figs.~\ref{fig:exploration_pbe} and \ref{fig:exploration_pm7}, respectively.

From Fig.~\ref{fig:exploration_pbe} it can be seen that the size of the training data set initially increases.
This is due to the low prediction confidence at the beginning of the exploration.
The data set increases until the prediction uncertainty is below $\sigma_\text{thresh}^2$ (shown as a horizontal line in Fig.~\ref{fig:exploration_pbe}, right).
This is the point at which the predictions made by the GP are trusted for the first time.
If, however, the exploration reaches regions of chemical space that are distant to the previously explored ones, the confidence will drop and new reference calculations will be required.
This can be observed in Fig.~\ref{fig:exploration_pbe}, right, where the variance exceeds $\sigma_\text{thresh}^2$.
Naturally, the total number of reference calculations for the entire exploration depends on the target and $\sigma_\text{thresh}^2$.
Finally, it can be seen that the backtracking mechanism described in Section~\ref{subsec:protocol} is indeed necessary.
In Fig.~\ref{fig:exploration_pbe}, for $\sigma_\text{thresh} = 6$ kJ/mol at $i = 651$,
the GP is updated and some predictions which previously were inside the confidence bound now lie outside of it.
Consequently, data points are added to the data set followed by an update of the GP until all predictions are within the confidence bound.

Fig.~\ref{fig:exploration_pm7} shows that a larger data set is required for the target $\Delta H^\text{G4MP2}_\text{PM7}$ to reach a
standard deviation of 15 kJ/mol than that for the target $\Delta H^\text{G4MP2}_\text{PBE}$ to reach a standard deviation of 8 kJ/mol.
This finding is in accordance with the results presented in Table~\ref{tab:at}.
Calculation of the enthalpy of atomization is faster with PM7 than that with PBE by about an order of magnitude (for the systems studied in this work).
However, because the exploration with PM7 as the base method requires far more computationally expensive G4MP2 reference calculations
(which take more than three orders of magnitude longer than PBE calculations for the systems studied in this work),
the overall exploration takes longer with PM7 than that with PBE as the base method.
We note that the time required for the evaluation of the kernel and GP predictions is negligible for data sets of this size.
This illustrates the philosophy of the $\Delta$-machine learning approach that should work more efficiently for the physically more reliable model
(in our case, this is PBE).
As a result, given a required confidence level, a trade-off needs to be found between the required number of reference calculations and the computational effort of the base method.
Results of an exploration starting from a different node are provided in the Supporting Information.

\section{Conclusions}

In this work, a novel approach for the rolling improvement of quantum chemical results through the application of GPs is presented.
By learning the error of an efficient quantum chemical method with respect to some reference method of higher accuracy,
we obtained accurate standard enthalpies of formation for configurational isomers of $\text{C}_7\text{H}_{10}\text{O}_2$ stoichiometry.
Accurate differences in standard enthalpy between isomers are accessible as well.
Furthermore, we showed that the uncertainty estimates provided by our predictive model for both the standard enthalpy of formation of molecules and
the difference in the standard enthalpy between molecules are reliable.
If the uncertainty associated with a particular calculation is above a given threshold,
then the chosen reference method will be employed to produce additional reference data.
In this way, reference calculations are performed only if truly necessary, i.e., if regions of chemical space unknown to our model are approached
and explored.
The approach presented in this work is independent of the chosen molecular descriptor and can also be carried out with more involved machine learning methods that provide error estimates.
In addition, we emphasize that our approach is independent of the chosen electronic structure methods, ranging from semiempirical and tight-binding models to
multiconfigurational approaches with multireference perturbation theory.
Through \textit{backtracking}, previous predictions are validated by the updated model to ensure that uncertainties remain within the given confidence bound.

Our approach will be beneficial for mechanism-exploration algorithms,\cite{Ohno2008,Maeda2013,Rappoport2014,Bergeler2015,Zimmerman2015,Habershon2016,Kim2018}
of which our \texttt{Chemoton}\cite{Simm2017a} algorithm is one example designed to be applicable to molecules from the whole periodic table of elements.
The combination with our \texttt{KiNetX}\cite{Proppe2018} algorithm for kinetic modeling under uncertainty propagation is currently being investigated in our laboratory.
In this way, reliable first-principles explorations of those portions of chemical reaction space that are relevant for a specific chemical problem will become accessible.
Obviously, this will require the accessibility of accurate reference calculations on demand.
For instance, our multiconfigurational diagnostic\cite{Stein2017} will allow one to decide on the
singlereference vs.\ multireference nature of the molecular structure subjected to a reference calculation.
For single-reference cases, explicitly correlated, local coupled cluster calculations\cite{Ma2018} are the method of choice
as they can be easily launched in an automated manner and are known to be highly accurate.
For multiconfigurational cases, automated complete active space-type calculations can be
launched with our fully automated procedure\cite{autoCASWeb2018} for the selection of active orbital spaces\cite{Stein2016,Stein2016a,Stein2017a}.

\section*{Computational Methodology}

The data set employed in this study is a subset of the GDB-17 data set.\cite{Ruddigkeit2012}
All G4 geometries were taken from ref.~\onlinecite{Ramakrishnan2014}.
The list of unique identifiers of the structures contained in this data set can be found in the Supporting Information.
G4MP2 enthalpies of atomization were also taken from ref.~\onlinecite{Ramakrishnan2014}.
DFT enthalpies of atomization were based on electronic energies obtained with
the PBE exchange-correlation functional\cite{Perdew1996a} and a double-$\zeta$ basis.\cite{Dunning1970}
DFT calculations were performed with the program packages \texttt{Q-Chem} (version 4.3).\cite{Shao2015}
Vibrational frequencies and rotational constants were taken from ref.~\onlinecite{Ramakrishnan2014}.
Accordingly, $\Delta H^\text{G4MP2}_\text{PBE}$ is given by the difference in G4MP2 and PBE electronic energies of atomization as the
nuclear contributions cancel in this setup.
By contrast, PM7 enthalpies of atomization were calculated from enthalpies of formation obtained with the MOPAC program (version 2016).\cite{MOPAC2016}

The SOAP average kernel was evaluated with the \texttt{glosim} package.\cite{De2016}
Following previous work,\cite{Bartok2013,Ferre2017} we chose an exponent of $\zeta = 4.0$.
In addition, we set the Gaussian width parameter to be $\sigma=0.3 \, \textup{\AA}$ and the cutoff radius to be $R_\text{cut}=4.0 \, \textup{\AA}$.
Furthermore, we chose the number of radial and angular functions to be 12 and 10, respectively.
Our model would likely benefit from an exhaustive search over hyperparameters; however, consistent with previous findings,\cite{De2016}
the performance of the kernel is not highly sensitive to the chosen set of parameters.

GP predictions were carried out with the library \texttt{GPy}.\cite{GPyWeb2016}
Data analysis and visualization were performed with the Python libraries \texttt{pandas}\cite{McKinney2010} and \texttt{matplotlib}\cite{Hunter2007}, respectively.
The graphical representation of the reaction network was created by the \texttt{Graphviz} program.\cite{Gansner2000}

\section*{Acknowledgments}

This work has been financially supported by the Schweizerischer Nationalfonds. 
G.N.S.\ gratefully acknowledges support by a Ph.D.\ fellowship of the Fonds der Chemischen Industrie.

%%%%%%%%%%%%%%%%%%%%%%%%%%%%%%%%%%%%%%%%%%%%%%%%%%%%
\section*{Supporting Information}

In the supporting information of the published article, 
the data set and the reaction network employed in this work can be found.
In addition, it contains a figure on the GPs' prediction accuracy.
Finally, the results of an additional exploration are provided.
This exploration differs in the initial node (i.e., the starting point of the exploration) from the one discussed in the main text.

%%%%%%%%%%%%%%%%%%%%%%%%%%%%%%%%%%%%%%%%%%%%%%%%%%%%

%\bibliographystyle{achemso2}
%\bibliography{references}
\providecommand{\refin}[1]{\\ \textbf{Referenced in:} #1}

\end{document}